**Title:** FreeDSM and the Gaia4Sustaniability project: a light pollution meter based on IoT technologies

**Introduction:**
Light pollution is a rapidly growing environmental problem, particularly in urbanized areas of developed countries. This phenomenon not only hinders astronomical observation, but also has a negative impact on ecosystems and human health, in addition to being an economic waste. To help to control its expansion, we introduce the Free Dark Sky Meter, a photometer based on Internet of Things (IoT) technologies that enables continuous and efficient monitoring at a low price and on an open data platform. This device seeks to provide a more affordable and accessible tool to measure light pollution.

**Description of the approach:**
FreeDSM (see Figure 1) is a simple and inexpensive device that was specifically designed to measure the amount of light in all sorts of environments with a DIY philosophy (https://gitlab.citic.udc.es/lia2-publico) . Its ubiquity is a crucial factor in effectively addressing light pollution. The device is based on an ESP32 microcontroller, known for its low cost and energy efficiency, and uses a light sensor to quantify the light intensity in the area. Additionally, it incorporates sensors to record temperature and humidity, allowing for a more complete analysis of the environment. FreeDSM is easily rechargeable with a battery, and offers the ability to be charged by any USB-C source or by a solar panel. Its open source design allows anyone to build and improve the device without licensing costs.

FreeDSM was calibrated with commercial instruments and a data-driven philosophy, using artificial neural networks and gradient descent techniques to adjust the coefficients for the calculation of the calibrated magnitudes per square arcsecond (MPSAS). Data from FreeDSM devices are collected and processed through a centralized system (https://g4s.citic.udc.es/), which aggregates the information from multiple devices to ensure comprehensive coverage and accuracy. This system allows for the effective management and analysis of the data collected, providing a robust foundation for further processing.

The system then uses a reference value for natural sky brightness to assess light pollution levels (Gambons model https://gaia4sustainability.eu/gambons/) (Masana, Carrasco, Bará, & Ribas, 2021). This reference value is calculated for the specific location and time of the measurement using a night sky brightness model and compared with the measured FreeDSM value. The Gambons model leverages data from the Gaia satellite, allowing for highly accurate real-time calculations of expected natural light levels, which are crucial for determining the true extent of light pollution. As shown in Figure 2, FreeDSM is calibrated alongside other commercial instruments to ensure accuracy and consistency.

**Concluding Paragraph:**
The development of FreeDSM represents a significant advance in light pollution monitoring because its low cost makes it accessible to the average user. Its ability to provide detailed real-time information on artificial light levels provides access to critical data that is essential for assessing the impact of light on ecosystems. Additionally, the integration of the Gambons model, which utilizes data from the Gaia satellite, significantly enhances the precision of light pollution measurements, offering a powerful tool for environmental monitoring. This information is fundamental to the design and implementation of informed mitigation strategies. As FreeDSM becomes more widely adopted, it is expected to facilitate better understanding and management of light pollution, thereby promoting more effective and sustainable management of the lighting environment. The FreeDSM device and the Gambons (Masana, Bará, Carrasco, & Ribas, 2022) model are part of the Gaia4Sustainability Project at https://gaia4sustainability.eu.

**Acknowledgements:**
This research was funded by the Spanish Ministry of Science MCIN / AEI / 10.13039 / 501100011033 and the European Union Next Generation programme EU/PRTR through the coordinated grant PDC2021-121059 (https://gaia4sustainability.eu, C21 University of Barcelona, and C22 University of A Coruña). Later supported by EU Horizon Europe [HORIZON-CL4-2023-SPACE-01-71] SPACIOUS 101135205, and the EU FEDER Galicia 2021-27 ED431G 2023/01.

Mario Casado Diez on behalf of the Gaia4Sustainability team.
CIGUS CITIC, Centre for Information and Communications Technologies Research, Universidade da Coruña, A Coruña, Spain
e-mail: mario.diez@udc.es

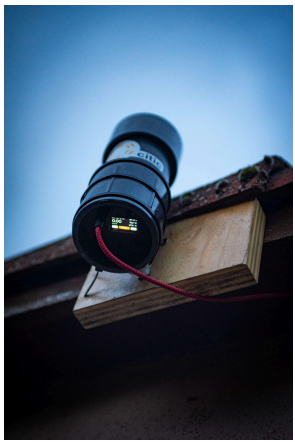

**Figure 1.** The FreeDSM device, as shown in the image, showing its design. Photograph taken by the author.

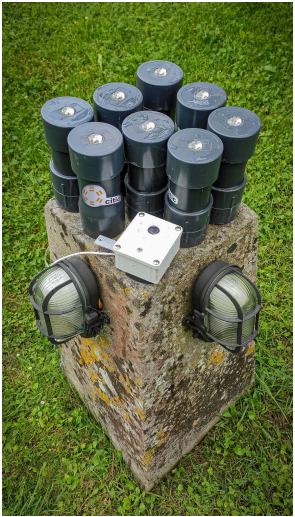

**Figure 2.** Calibration setup of FreeDSM with other commercial instruments (TESS). Photograph taken by the author.

**References:**

Masana, E., Carrasco, J. M., Bará, S., & Ribas, S. J. (2021). A multi-band map of the natural night sky brightness including Gaia and Hipparcos integrated starlight. Monthly Notices of the Royal Astronomical Society, 501, 5443. https://arxiv.org/abs/2101.01500

Masana, E., Bará, S., Carrasco, J. M., & Ribas, S. J. (2022). An enhanced version of the Gaia map of the brightness of the natural sky. International Journal of Sustainable Lighting, 24, 1-12.